\documentstyle[preprint,floats,aps,epsf,pre]{revtex}
\begin{document}
\draft

\title{Relation between Stochastic Resonance and
Synchronization of Passages in a Double-Well System}
\author{Mangal C. Mahato and A.M. Jayannavar}
\address{Institute of Physics, Sachivalaya Marg,
Bhubaneswar-751005, India}

\maketitle

\begin{abstract}

We calculate, numerically, the residence times (and their
distribution) of a Brownian particle in a two-well system under
the action of a periodic, saw-tooth type, external field. We
define hysteresis in the system. The hysteresis loop area is
shown to be a good measure of synchronization of passages
from one well to the other. We establish connection between this
stochastic synchronization and stochastic resonance in the
system.
\end{abstract}

\pacs{PACS numbers: 82.20.Mj, 05.40.+j, 75.60.Ej}

Nature, presumably, exploits the phenomena of stochastic
resonance (SR) to its advantage\cite{Moss,Doug} to tune in to a
desired signal. However, its discovery, initially as a
theoretical invention to explain recurrence of ice
ages\cite{Benz,MW}, is barely a decade and half old.  SR is a
nonlinear phenomena wherein an input noise is partially
rectified by the system to obtain enhanced output signal at the
input signal frequency. It is reflected as a peak in the output
signal-to-noise ratio (SNR) as a function of input noise
strength. The nature of output signal in the nonlinear system, 
however, depends on the combined effect of the input signal and
the input noise.  Since the input signal, in general, is often
considered to be weak (subcritical), in the absence of noise no
output signal is obtained. The output signal considered is an
averaged effect of a large number of realizations of fluctuating
forces (noise) in combination with the input signal. The output
signal, therefore, may not have the same form as the input
signal. However, one expects the output signal to have its
dominant time scales of variation nearly synchronous with those
of the input signal. The degree of synchronization, however,
will depend on the noise strength as well as the input signal
parameters. It is, therefore, interesting to study how the
output signal gets synchronized with the input signal, and, in
particular, whether SR occurs exactly when there is maximum
synchronization between the two signals. In this work we show,
by calculating the residence time distributions in a
periodically forced two-well potential, that there is a close
connection between these two phenomena. 

We define hysteresis in a two-well system, as explained below,
such that its loop area is a good measure of synchronization of
passages from one well to the other under the influence of a
Gaussian white noise $\hat f (t)$ when subjected to an external
periodic field $h(t)$. Just in order to make the explanation
simpler and more transparent we define hysteresis loop from the
distribution of first-passages from one well to the other. ( The
explanation may subsequently be simply carried through to the
main part of this work wherein residence time distribution is
used.)  The procedure is illustrated in Figures 1(a-d). $\rho
(\tau)$ is the distribution of first-passage-times, $\tau$, from
one well to the other (Fig. 1b) obtained by solving the
overdamped Langevin equation
\begin{equation}
\dot m (t)=-\frac{\partial U(m)}{\partial m}+{\hat f(t)}
\end{equation}
numerically (for a large number of realisations), where
\begin{equation}
U(m)=-\frac{a}{2}m^2 +\frac{b}{4}m^4 -mh(t)
\end{equation}
represents the two-well potential in the presence of an external
field $h(t)$ which is periodic in time $t$. The fluctuating
forces $\hat f(t)$ satisfy:$\left< \hat f(t) \right>=0$ and
$\left< \hat f(t)\hat f(t') \right>=2D\delta (t-t')$, where $D$
is the strength of the noise. The sequence of $\tau$'s is then
used to calculate the distribution $\rho (h(\tau))$ (Fig. 1c) of
field values $h(\tau)$ at which first-passages occur from the 
right well to the left one. From $\rho (h(\tau ))$ we obtain the
upper half of the hysteresis loop $M(h)$ (Fig. 1d),
\begin{equation}
\frac{M(h)}{h_c}=1-2 \int_{h}^{h_0} \rho (h') dh',
\end{equation}
the other half being obtained by symmetry\cite{Maha}. Here $h_0$ is the
amplitude and $\vert h_c \vert$ is the minimum value of the
field $h(t)$ (Fig. 1a) beyond which one of the two wells of
$U(m)$ disappears. 

From Fig. 1b we see that the peaks of $\rho (\tau )$ occur
periodically and centered each time around which $h(t) = -h_0$
at which the potential barrier of passage from the right to the
left well is the least. Now, perfect synchronization of passages
would mean sharp ($\delta $-function) periodic peaks in $\rho
(\tau )$ with periodicity $T_0$ of $h(t)$ so that just one sharp
peak appears at $h=-h_0$ in $\rho (h(\tau ))$. This case would
yield a rectangular hysteresis loop and hence with the largest
possible area. On the other hand, if the passages take place all
over randomly, for the other extreme case of least
synchronization, $\rho (h(\tau ))$ would be uniform resulting in
a hysteresis loop of zero area. The hysteresis loop area, thus,
provides a measure of the degree of synchronization of passages.
It is to be noted that the calculation of hysteresis loop
automatically takes into account of informations buried in {\it
{all}} the peaks of the passage-time distribution $\rho (\tau)$.
It has earlier been shown\cite{Maha} that hysteresis loop area
so calculated acquires a maximum as a function of the input
noise strength $D$ as well as a function of the sweep rate
$\vert \dot h \vert $ (equivalent to the frequency) of the
saw-tooth type periodic input signal. We now discuss our
numerical experiment wherein the above explanation about the 
measure of synchronization should carry through, though not in
as obvious a manner, from the residence time distributions in
each of the two wells. 

In order to obtain the residence time distributions, $\rho _1
(\tau )$ and $\rho_2 (\tau)$ in each of the wells 1 and 2,
respectively, we monitor the trajectory $m(t)$ of the particle
[Eqs. (1) and (2)] for a long time and keep putting markers on
the time axis whenever a passage from one well to the other
takes place. We take $h(t)$ of the same form as in Fig. 1d but
with $h(t=0)=0$. We consider the passage to take place only if
the trajectory $m(t)$ crosses the inflexion point on the far
side of the maximum of the potential barrier separating the two
wells. From the markers on the time axis we also obtain
the jump field values $h(t)$ for switching from one well to the
other and hence the corresponding distributions
$\rho_{12}(h(t))$ and $\rho_{21}(h(t))$. Figures 2a and 2b show
the typical plots of $\rho_1(\tau)$ and $\rho_{21} (h)$,
respectively. Notice that we have, now, calculated
$\rho_{12}(h)$, etc., properly so that the distributions spread 
over the entire period of $h(t)$, and includes both ascending as
well as descending parts of $h(t)$. It is now easy to calculate
the probability $m_2(h)$ that the trajectory lies in the well 2 
when the field value $h(t)=h$, from the discrete equation,
\begin{equation}
m_2(h)=m_2(h-\Delta h)-m_2(h-\Delta h) \rho_{21}(h)\Delta h
+m_1(h-\Delta h)\rho_{12}(h)\Delta h,
\end{equation}
with negligibly small $\Delta h$, and similarly for
$m_1(h)=1-m_2(h)$. 
We, then, calculate the hysteresis loop $m(h)=m_2(h)-m_1(h)$
iteratively that satisfies the closed loop condition
$m(h(t+T_0))=m(h(t))$. Figure 2c shows a typical (asymptotically 
stationary) hysteresis loop. Please note that the
hysteresis loop need not be saturated for every case in contrast
to Fig. 1d. We, now, discuss the results on the stochastic
synchronization(SS) of passages from the variation of hysteresis
loop area as a function of noise strength $D$ and also as a
function of sweep rate $\vert \dot h \vert $.

Throughout our calculation, we take $a=2.0$ and $b=1.0$ in the
expression for $U(m)$, so that the barrier height when $h(t)=0$
is 1. We take $h_0 = .7h_c$ and $.9h_c$ and calculate for each
of these cases, the distributions $\rho_1(t)$ and $\rho_2(t)$
and also the passage field distributions $\rho_{12}(h(t))$ and
$\rho_{21}(h(t))$ for various values of $\dot h$. The results
obtained for both the field amplitudes $0.7h_c$ and $0.9h_c$
have qualitatively similar trends. Figure 3 shows the variation
of hysteresis loop area as a function of noise strength $D$. We
observe that the hysteresis loop area initially increases as D
is increased from a small value, attains a maximum value at
$D=D_{ma}$, say, and then decreases gradually as $D$ is
increased further. It shows that at $D=D_{ma}$ the output signal
is most synchronized with the input signal $h(t)$. The
hysteresis loop area is also found to show maximum as a function
of $\dot h$ (Fig. 3b).  

From the residence time distributions we calculate the Fourier
transform and each of the components are squared to obtain the
power spectral density\cite{Pres}. As expected large peaks are
obtained at regular intervals. We calculate the ratio of the
height of the first peak and the (background) noise level at the
same frequency. This signal-to-noise ratio (SNR) is plotted in
Fig. 4 as a function of $D$. The errors in calculating the SNR
are large mostly because of the arbitrariness in fixing the
(low) background noise level. However, the errors do not affect the
general trends of our results.  We, indeed, find SR in the usual
sense of the SNR maxima. Interestingly, the value of
$D=D_{SNR}(\dot h)$ at which SR occurs, is not the same as the
value of $D=D_{ma}(\dot h)$ at which the passages are the most
synchronized. However, $D_{SNR}(\dot h)$ and $D_{ma}(\dot h)$
are quite close and they tend to become closer as $\dot h
\rightarrow 0$, as shown in Fig. 5. (A finer examination to
smaller $\dot h$ is beyond the computing power available to us.)
SR and SS are, thus, $\it not$ unrelated\cite{Gamm,Buls}. SR may
,therefore, be a natural manifestation of SS. Moreover, SR,
defined as the maximum of SNR appears not only as a function of
noise strength $D$ but {\it also} as a function of the sweep
rate $\dot h$ (or equivalently the frequency) of the external
field (input signal) as shown in Fig. 6 ({\it cf.} with Fig.
3b). 

In conclusion we state that hysteresis loop area, which is an
average effect of all the peaks in the residence time
distribution, is a good measure of synchronization of passages
with the input signal. In an earlier work\cite{Gamm}, by taking
the only first few peaks, and in particular the first peak, of the
residence time distribution, SR was given an alternative description
and was shown to be a bonafide resonance. On closer examination we
find that rest of the peaks too play important role. We, however,
concretise their assertion, by taking into account {\it all} the
peaks of the residence time distribution through the hysteresis loop
area, that SR is a genuine resonance, at least in a double-well
potential system, and is due to the synchronized response of the
system to the input periodic signal. Also, the degree of
synchronization of passages and the SNR show maximum as a
function of the sweep rate (or equivalently the frequency) of
the input signal\cite{BG}.

\begin{figure}

\caption{Shows (a) the time variation of the external
field $h(t)$, (b) the first-passage-time distribution $\rho
(\tau)$, (c) the passage field distribution $\rho (h)$, and (d)
the corresponding hysteresis loop calculated from $\rho(h)$, for
$h_0=0.7h_c$, $D=0.3$, and the period of $h(t)$, $T_0=28.0$.}
~~~~

\caption{Shows (a) the residence time distribution $\rho
_1(\tau)$ in the well 1, (b) the passage field distribution $\rho
_{21}(h)$ for passage from well 2 to well 1, and (c) the
corresponding hysteresis loop $m(h)$ for $h_0=0.9h_c$, $D=0.2$,
and period of $h(t)$, $T_0=36.0$.}
~~~~

\caption{Plots of (a) hysteresis loop area $A$ as a
function of $D$ for $\dot h=0.05h_c (\circ)$, $0.1h_c (\Box)$,
$0.2h_c (\diamond)$, $0.4h_c (\bigtriangleup)$, and $0.6h_c 
(\bigtriangledown)$, and (b) hysteresis loop area $A$ as a
function of field sweep rate $\dot h$ for $D=0.1 (\circ)$, $0.15
(\Box)$, $0.2 (\diamond)$, and $0.3 (\bigtriangleup)$ for
$h_0=0.9h_c$. In this and in the rest of the figures the lines
joining the points are only to guide the eye.}
~~~~

\caption{Plots of signal-to-noise ratio (SNR) as a
function of $D$, for $\dot h/h_c=0.05 (\circ)$, $0.1 (\Box)$,
$0.2 (\diamond)$, $0.4 (\bigtriangleup)$, $0.5 (\triangleleft)$,
$0.6 (\bigtriangledown)$, and $0.72 (\triangleright)$, for
$h_0=0.9h_c$.} 
~~~~

\caption{Plots the peak positions of the plots of
hysteresis loop area $A$ versus $D (\circ)$ and those of SNR
versus $D (\Box)$, for $h_0=0.7h_c$ (empty symbols, solid
joining lines) and for $h_0=0.9h_c$ (filled symbols, dashed
joining lines).} 
~~~~

\caption{Shows SNR as a function of $\dot h$ for
$h_0=0.9h_c$, for various values of $D=0.1 (\circ)$, $0.15
(\Box)$, and $0.2 (\diamond)$.}
~~~~
\end{figure}
~~~~~
\end{document}